\documentclass[12pt]{article}
\usepackage{amsmath,amssymb,amsfonts}
\newtheorem{theorem}{Theorem}
\newtheorem{prop}[theorem]{Proposition}

\newtheorem{exe}[theorem]{Exercise}
\newtheorem{exa}[theorem]{Example}

\newtheorem{remark}[theorem]{Remark}

\newenvironment{proof}[1][Proof]{\noindent\textbf{#1.} }{\null\hfill
\rule{0.5em}{0.5em}}
\newcommand{\di}{\mathrm{d}}
\newcommand{\la}{\lambda}
\newcommand{\tr}{\mathrm{tr}\,}

\newcommand{\CS}{{\cal S}}

\newcommand{\CN}{{\cal N}}

\newcommand{\CM}{{\cal M}}
\newcommand{\CQ}{{\cal Q}}

\newcommand{\RR}{{\mathbb R}}

\def\al{\alpha}

\def\la{\lambda}

\newcommand{\rref}[1]{(\ref{#1})} 

\def\bih{bi-Ham\-il\-tonian}
\def\varb{\bih\ manifold}

\begin{document}

\begin{center}
{\Large \bf A 3-component extension\\ \smallskip
of the Camassa-Holm hierarchy}
\end{center}
\vspace{0.5truecm}
\begin{center}
Laura Fontanelli, Paolo Lorenzoni\\ \medskip
Dipartimento di Matematica e Applicazioni\\
Universit\`a di Milano-Bicocca\\
Via Roberto Cozzi 53, I-20125 Milano, Italy\\
laura.fontanelli@unimib.it, paolo.lorenzoni@unimib.it\\ \bigskip
Marco Pedroni\\ \medskip
Dipartimento di Ingegneria Gestionale e dell'Informazione\\
Universit\`a di Bergamo\\
Viale Marconi 5, I-24044 Dalmine (BG), Italy\\
marco.pedroni@unibg.it
\end{center}
\begin{abstract}
We introduce a bi-Hamiltonian hierarchy on the loop-algebra of $\mathfrak{sl}(2)$ endowed with a suitable Poisson pair. It gives rise to the usual CH hierarchy by means of a bi-Hamiltonian reduction, and its first nontrivial
flow provides a 3-component extension of the CH equation.
\end{abstract}
{\footnotesize
~
}

\section{Introduction}

One of the many remarkable facts concerning the theory of integrable PDEs is that almost all these equations can be obtained as suitable reductions from (integrable) hierarchies living on
loop-algebras. The main example is the Drinfeld-Sokolov reduction \cite{DS}, leading from (the loop-algebra of) a simple Lie algebra $\mathfrak{g}$ to the so-called generalized Korteweg-deVries (KdV) equations. It is well-known (see, e.g., \cite{Dickeybook})
that the choice
$\mathfrak{g}=\mathfrak{sl}(n)$ gives rise to the Gelfand-Dickey hierarchies. In particular,
$\mathfrak{sl}(2)$ corresponds to the KdV hierarchy, whose most famous member is the KdV equation
\begin{equation}
u_{t}=-3uu_{x}+u_{xxx}\ ,
\end{equation}
describing shallow water waves. Another important model for such waves is the
Camassa-Holm (CH) equation
\begin{equation}
\label{cheq}
u_{t}-u_{txx}=-3uu_{x}+2u_{x}u_{xx}+uu_{xxx}\ ,
\end{equation}
derived and studied in \cite{CH}, and later found to be in the class of integrable systems introduced
in \cite{FoFu}. Like KdV, it belongs to a hierarchy of commuting
evolution equations and possesses a bi-Hamiltonian formulation. Also because of the existence of travelling wave solutions with discontinuous first derivative (the famous {\em peakons\/}), the CH equation is presently one of the most studied examples of integrable PDEs. 
Many papers (see, e.g., \cite{BSS1,BSS2,KM,McK}) have investigated its connections with the KdV hierarchy, and the Whitham modulation theory for CH has been recently discussed in \cite{AbGr}. Despite the importance of the CH hierarchy, it was not known if it shares with the other integrable PDEs the property mentioned at the beginning of this introduction, namely, if it can be seen as the reduction of a matrix hierarchy. The first step to solve this problem was made in \cite{LoPe}, where it has been shown that the bi-Hamiltonian structure of CH is the 
reduction of a suitable bi-Hamiltonian structure on the space $\CM=C^\infty(S^1,\mathfrak{sl}(2))$
of $C^{\infty}$ maps from the unit circle to $\mathfrak{sl}(2)$. The reduction process --- called bi-Hamiltonian reduction ---
is a particular instance of the Marsden-Ratiu reduction \cite{MR} and can be canonically performed on every bi-Hamiltonian manifold. In \cite{CMP} it has been introduced and applied to the KdV hierarchy, while in \cite{CP,P} its equivalence with the Drinfeld-Sokolov reduction has been showed.

In this paper we make the second (and final) step, showing that there exists a bi-Hamiltonian hierarchy on $\CM$ --- to be called the {\em matrix CH hierarchy\/} --- that gives rise to the usual (i.e., scalar) CH hierarchy after
that the bi-Hamiltonian reduction is performed. This allows us to interpret also CH as a reduced hierarchy, and to define a 3-component extension of the CH equation. This should be related to the 2-component extension introduced in \cite{LZ} in the framework of the Dubrovin-Zhang theory of deformations of bi-Hamiltonian structures of hydrodynamic type, and further studied in more details in \cite{CLZ,F}.

The paper is organized as follows: In Section 2 we recall the bi-Hamiltonian reduction process and its application to the CH case. In Section 3 we present the 3-component extension of the CH equation, then we show that it belongs to the
(nonlocal) matrix CH hierarchy, studied in Section 4. Section 5 is devoted to conclusions, including a brief discussion on the local matrix CH hierarchy.

\par\smallskip\noindent
{\bf Acknowledgments.}
We wish to thank Gregorio Falqui, Franco Magri, and Giovanni Ortenzi
for useful discussions and suggestions.
This work has been partially supported by INdAM-GNFM under the
research project {\em Onde nonlineari, struttura tau e geometria
delle variet\`a invarianti: il caso della gerarchia di Camassa-Holm\/}, and
by the European Community through the FP6 Marie Curie RTN {\em ENIGMA}
(Contract number MRTN-CT-2004-5652).
 M.P. would like to thank for the hospitality the Department
 {\em Matematica e Applicazioni\/} of the Milano-Bicocca
 University, where most of this work was done.

\section{Some information about the bi-Hamiltonian reduction}

In this section we review some facts on the geometry of bi-Hamiltonian manifolds (see, e.g., \cite{pondi,CFMP3}), and we recall from \cite{LoPe} that the bi-Hamiltonian structure of the (usual) CH hierarchy can be obtained by means of a reduction.

Let $(\CM,P_1,P_2)$ be a \varb, i.e., a manifold $\CM$ endowed with two
 compatible Poisson tensors $P_1$ and
$P_2$.
 Let us fix a symplectic leaf $\CS$ of $P_1$ and consider the distribution
$D=P_2(\mbox{Ker}P_1)$ on $\CM$.

\begin{theorem} The distribution $D$ is integrable. If $E=D\cap T\CS$ is the distribution induced by $D$ on $\CS$ and the quotient space $\CN=\CS/E$ is a manifold, then it is a \varb.
\end{theorem}

Whenever an explicit description of the quotient manifold $\CN$ is not
 available, the following technique to compute the reduced \bih\
 structure (already employed in \cite{CP} for the Drinfeld-Sokolov case)
 is very useful. Suppose $\CQ$ to be a submanifold of $\CS$ which is trasversal to the distribution $E$, in the sense that
\begin{equation}
\label{split}
T_p\CQ\oplus E_p=T_p\CS\qquad\mbox{for all $p\in\CQ$}\ .
\end{equation}
Then $\CQ$ (which is locally diffeomorphic to $\CN$) also inherits a \bih\ structure from $\CM$. The reduced Poisson pair on $\CQ$ is given by
\begin{equation}
\label{redP}
\left(P_i^{\mbox{\scriptsize rd}}\right)_{p}\al=\Pi_p\left((P_i)_p\tilde\al\right)\ ,\qquad i=1,2\ ,
\end{equation}
where $p\in\CQ$, $\al\in T^*_p\CQ$, $\Pi_p:T_p\CS\to T_p\CQ$ is the projection relative to \rref{split}, and $\tilde\al\in T^*_p\CM$ satisfies
\begin{equation}
\tilde\al|_{D_p}=0\ ,\qquad \tilde\al|_{T_p\CQ}=\al\ .
\end{equation}

Let us suppose now that $\{H_j\}_{j\in\mathbb Z}$ be a bi-Hamiltonian hierarchy on $\CM$, that is,
$P_2 dH_j=P_1 dH_{j-1}$ for all $j$. This amounts to saying that
$H(\lambda)=\sum_{j\in\mathbb Z}H_j\lambda^{-j}$ is a (formal) Casimir of the {\em Poisson pencil\/} $P_1-\lambda P_2$. The bi-Hamiltonian vector fields associated with the hierarchy can be reduced on the quotient manifold $\CN$ according to
\begin{prop}
The restrictions of the functions $H_j$ to $\CS$ are constant along the distribution
$E$, and therefore they give rise to functions on $\CN$.
These functions form a bi-Hamiltonian hierarchy with respect to the reduced Poisson pair. The vector fields $X_j=P_2dH_{j}=P_1dH_{j-1}$ are tangent to
$\CS$ and project on $\CN$. Their projections are the vector fields associated with the reduced hierarchy.
\end{prop}

Now let $\CM = C^{\infty}(S^1,\mathfrak{sl}(2))$
 be the loop-space on the Lie algebra of traceless $2\times 2$ real matrices.
The tangent space $T_S\CM$ at $S\in \CM$  is identified with $\CM$ itself,
 and we will assume that $T_S\CM \simeq T^*_S\CM $ by the nondegenerate form
$$ \langle V_1, V_2\rangle = \int\mathrm{tr} (V_1(x)V_2(x))\, \di x,
 \qquad V_1,V_2 \in \CM\ , $$
where the integral is taken (here and in the rest of the paper) on $S^1$. 
It is well-known (see, e.g., \cite{LM}) that the manifold $\CM$ has a $3$-parameter family of compatible Poisson tensors
\begin{equation}
\label{pencil}
 P_{(\lambda_1, \lambda_2, \lambda_3)}=\lambda_1 \partial_x
+\lambda_2[\,\cdot\, ,S] +\lambda_3 [\,\cdot\,
, A]\ ,
\end{equation}
where $\lambda_1, \lambda_2, \lambda_3 \in \RR$ ,
 $A$ is a constant matrix in $\mathfrak{sl}(2)$, and
\begin{equation*}
S=\begin{pmatrix} p & q\\ r & -p\end{pmatrix}\in\CM\ .
\end{equation*}
In this paper we are interested in the pencil
\begin{equation}
\label{poipen} P_{\la}=P_1-\lambda P_2=[\,\cdot\, ,S] -\lambda
(\partial_x  +[\,\cdot\, , A])\ ,
\end{equation}
obtained from (\ref{pencil}) setting $\lambda_2=1$, $\lambda_3=\lambda_1=-\lambda$, and
$$A=\frac12\left( \begin{array}{cc}
0&1\\
1&0
\end{array}\right)\ . $$
In \cite{LoPe} the bi-Hamiltonian reduction procedure has been applied to
 the pair $(P_1,P_2)$. In this case
\begin{equation*}
D_S=\left\{\begin{pmatrix}
(\mu p)_x+\frac{1}{2}(q-r)\mu & (\mu q)_x+2\mu p \\
(\mu r)_x-\mu p & -(\mu p)_x-\frac{1}{2}(q-r)\mu
\end{pmatrix}
\mid \mu\in C^\infty(S^1,\mathbb{R})\right\}\ .
\end{equation*}
The distribution $D$ is not tangent to the generic symplectic leaf of
 $P_1$, but it is  tangent to the symplectic leaf
\begin{equation}
\CS=\left\{\begin{pmatrix} p & q \\ r & -p \end{pmatrix}
\mid p^2+qr=0, (p,q,r)\not=(0,0,0)\right\}\ ,
\end{equation}
so that $E_p=D_p\cap T_p\CS$ coincides with $D_p$ for all $p\in\CS$.
It is not difficult to prove that the submanifold
\begin{equation}
\CQ=\left\{S(q)=\begin{pmatrix} 0 & q \\ 0 & 0 \end{pmatrix}
\mid q\in C^\infty(S^1,\mathbb{R}), q(x)\ne 0\ \forall x\in S^1\right\}
\end{equation}
of $\CS$ is transversal to the distribution $E$ and that the projection
$\Pi_{S(q)}:T_{S(q)}\CS\to T_{S(q)}\CQ$ is given by
\begin{equation}
\label{proj}
\Pi_{S(q)}:(\dot p,\dot q)\mapsto (0,\dot q-2\dot p_x)\ .
\end{equation}
The reduced bi-Hamiltonian structure (\ref{redP}) coincides with the bi-Hamiltonian structure
of the Camassa-Holm hierarchy (see \cite{LoPe} for details):
\begin{eqnarray*}
&&\left(P_1^{\mbox{\scriptsize rd}}\right)_{q} =2(2q\partial_x+q_x)\\
&&\left(P_2^{\mbox{\scriptsize rd}}\right)_{q} =2( -\partial_x^3+\partial_x)\ .
\end{eqnarray*}
Starting from the Casimir $\int\sqrt{q}\, \di x$ of $P_1^{\mbox{\scriptsize rd}}$ one constructs the local (or negative) CH hierarchy. The Casimir $\int q\, \di x$ of
$P_2^{\mbox{\scriptsize rd}}$ gives rise to the nonlocal (or positive) CH hierarchy, whose second flow is the CH equation \rref{cheq}. We refer to \cite{clop} and the references cited therein for more details, and for a discussion about a ``KP extension'' of the local hierarchy.

\section{A 3-component extension of the Camassa-Holm equation}

In this section we start to construct a bi-Hamiltonian hierarchy associated to the pencil
(\ref{poipen}). We consider the functional
\begin{equation}
\label{casp2}
H_1=-\frac{1}{2}\int (r+q)\,\di x\ ,
\end{equation}
which is easily seen to be a Casimir of $P_2$. Applying $P_1$ to the differential
$dH_1$ we obtain the vector field
\begin{equation}
\label{nlch1}
\begin{cases}
&\dot{p}=-\frac{1}{2}(r-q)\\
&\dot{q}=p\\
&\dot{r}=-p
\end{cases}
\end{equation}
This vector field is Hamiltonian also with respect to $P_2$, i.e., it can be written as 
$P_2 dH_2$, where 
\begin{equation}
\label{h2onm}
H_2[p,q,r]=\int(p^2-\gamma^2-\gamma_x^2)\,\di x\ ,
\end{equation}
where $\gamma$ satisfies the equation
\begin{equation}
\label{gamma}
\gamma_{xx}-\gamma=\frac{1}{2}(q-r)-p_x\ .
\end{equation}
To obtain this expression of $H_2$, let us start from
\begin{equation}
\label{dh2}
(dH_2)_x+[dH_2,A]=[dH_1,S]\ .
\end{equation}
and let us denote the components of $dH_2$ in the following way:
\begin{equation*}
dH_2=\begin{pmatrix} \alpha & \beta\\ \gamma & -\alpha\end{pmatrix}\ .
\end{equation*}
Equation (\ref{dh2}) in componentwise form reads
\begin{equation*}
\begin{cases}
&\alpha_x-\frac{1}{2}(\gamma-\beta+r-q)=0\\
&\beta_x-p+\alpha=0\\
&\gamma_x+p-\alpha=0
\end{cases}
\end{equation*}
The last two equations implies $(\beta+\gamma)_x=0$. If we suppose
 that $\beta=-\gamma$, then we have
\begin{equation*}
\begin{cases}
&\alpha=p+\gamma_x\\
&\beta=-\gamma\\
&\gamma_{xx}-\gamma=\frac{1}{2}(q-r)-p_x
\end{cases}
\end{equation*}
The expression of $H_2$ can be easily obtained evaluating $dH_2$ on a tangent vector $\dot S$.
 Indeed, we have
\begin{eqnarray*}
\langle dH_2,\dot{S} \rangle &=& \int[2\dot{p}(\gamma_x+p)-\gamma\dot{r}
+\gamma\dot{q}] \di x=\frac{\di}{\di t}\int p^2 \di x +\int\gamma(-2\dot{p}_x-\dot{r}
+\dot{q}) \di x\\
&=&\frac{\di}{\di t}\int p^2 \di x+2\int\gamma(\dot{\gamma}_{xx}-\dot{\gamma})\di x=
\frac{\di}{\di t}\int (p^2-\gamma^2) \di x-2\int\dot{\gamma}\dot{\gamma}_x \di x\\
&=&\frac{\di}{\di t}\int (p^2-\gamma^2-\gamma_x^2) \di x\ ,
\end{eqnarray*}
leading to \rref{h2onm}.

Hence the second vector field $P_1 dH_2$ of the hierarchy is given by
\begin{equation}
\label{nlch2}
\begin{cases}
&\dot{p}=-\gamma(r+q)\\
&\dot{q}=2q(\gamma_x+p)+2p\gamma\\
&\dot{r}=2p\gamma-2r(p+\gamma_x)
\end{cases}
\end{equation}
where $\gamma$ satisfies equation \rref{gamma}.

Let us show that the vector fields \rref{nlch1} and \rref{nlch2} project on the first members of the usual CH hierarchy. The reduction procedure for $P_1dH_1$ goes as follows:

- we restrict $P_1dH_1$ on the symplectic leaf $\CS$ and in particular
 on the points $(p=r=0)$ of the transversal submanifold $\CQ$;

- we project the restricted vector field
 according to the formula  (\ref{proj}).
\\
We obtain
\begin{equation*}
q_{t_1}=\dot{q}-2\dot{p}_x=-q_x
\end{equation*}
As far as $P_1 dH_2$ is concerned, its restriction at the points of $\CQ$  is
\begin{equation}
\label{nlch2onq}
\begin{cases}
&\dot{p}=-\gamma q\\
&\dot{q}=2q\gamma_x\\
&\dot{r}=0
\end{cases}
\end{equation}
where, from \rref{gamma}, one has that $\gamma_{xx}-\gamma=\frac{1}{2}q$. This shows that the usual change of dependent variable for the CH equation arises naturally in the reduction 
procedure. The projection of \rref{nlch2onq} on $T\CQ$ is
\begin{equation*}
q_{t_2}=\dot{q}-2\dot{p}_x=4q\gamma_x+2q_x\gamma\ ,
\end{equation*}
that is,
\begin{equation*}
\gamma_{xxt_2}-\gamma_{t_2}=4\gamma_{xx}\gamma_x-6\gamma\gamma_x+2\gamma\gamma_{xxx}\ ,
\end{equation*}
which becomes the Camassa-Holm equation \rref{cheq} after putting $u=-2\gamma$. 
Thus \rref{nlch2} is a 3-component extension of the Camassa-Holm equation.
Notice that the reduced Hamiltonian
\begin{equation*}
H_2[0,q,0]=-\int(\gamma^2+\gamma_x^2)\di x
\end{equation*}
is the first Hamiltonian of the Camassa-Holm equation.

Now we want to show that this procedure can be iterated, i.e., that the vector fields \rref{nlch1} and \rref{nlch2} belong to a (bi-Hamiltonian) hierarchy of (commuting) vector fields, whose reduction is the (scalar, nonlocal) CH hierarchy. To this aim, we construct a Casimir of the pencil
\rref{poipen} using the classical method of dressing transformations \cite{ZS,DS,CMP2}.

First of all we observe that the elements
$$V=\left( \begin{array}{cc}
v_1&v_2\\
v_3&-v_1
\end{array}\right) $$
 of $\ker{P_{\lambda}}$ satisfy the equation
\begin{equation}
\label{mateq}
-\la V_x + [V, S- \la  A] =0\ ,
\end{equation}
which implies that
\begin{equation*}
\frac{\di}{\di x} \tr V^2 = 0\ ,
\end{equation*}
so that the spectrum of $V$ does not depend on $x$:
\begin{equation} \label{tr}
  \tr \frac{V^2}{2} = F(\lambda)\ .
\end{equation}
Therefore, there exists a nonsingular matrix $K$ such that
\begin{equation*}
V(\lambda)=K\Lambda K^{-1}\ ,
\end{equation*}
where $$\Lambda = \left( \begin{array}{cc}
0&F(\lambda)\\ 1 & 0 \end{array}\right)\ .
$$
\begin{prop}
If $F(\lambda)$ does not depend on the point $S$, then $V(\la)$ is an exact 1-form whose potential is given by
\begin{equation}
\label{H}
H(\la)= \int \tr (M \Lambda)\, \di x\ ,
\end{equation}
where
\begin{equation}
\label{M}
 M= K^{-1}( S-\lambda A)K+\lambda K^{-1}K_x\ .
\end{equation}
\end{prop}
\begin{proof}
From (\ref{mateq}) it follows that
$$ -\lambda K^{-1}V_x K+ K^{-1}[V,  S - \lambda A] K=
-\lambda \Lambda_x+ [\Lambda, M] =[\Lambda, M]=0\ ,
$$
which implies that
 $\int \tr ( [M, K^{-1}\dot{K}] \Lambda)\, \di x=0$. Thus, for every tangent vector
$\dot S$, we have
\begin{eqnarray*}
\langle \di H, \dot{S} \rangle & = & \int \tr (\dot{M} \Lambda) \, \di x \,  = \,
\int \tr (K^{-1}\dot{S}K \Lambda) + \tr ( [M, K^{-1}\dot{K}] \Lambda) \, \di x \\
 & = & \int   \tr (\dot{S} K \Lambda K^{-1})\, \di x
= \int \tr (\dot{S} V)\, \di x =\langle  V,\dot{S} \rangle \ .
 \end{eqnarray*}
\end{proof}

Let us compute explictly $M$ and $H$. A possible choice for $K$ is
\begin{equation*}
K=\left( \begin{array}{cc}
v^{-\frac{1}{2}}& v_1v^{-\frac{1}{2}} \\ 0 & v^{\frac{1}{2}} \end{array}\right)\ ,
\end{equation*}
where $v=v_3$. From (\ref{M}) and (\ref{H}) we easily obtain
$$
M = \left( \begin{array}{cc}
0  & v^{-1} ( r-\frac{\lambda}{2}) F(\la)\\
v^{-1} ( r-\frac{\lambda}{2}) &0
\end{array}\right) \, = \,\frac{2r-\la}{2v} \Lambda
$$
and
\begin{equation}
H (\la) = \int \frac{  F(\la)(2 r- \la)}{v} \,\di x\ .
\end{equation}
The functional $H(\la)$ can also be written as
\begin{equation}
\label{Hla}
H (\la) = 2\la \sqrt{F(\la)} \int h \,\di x\ ,
\end{equation}
where the density $h$ is clearly defined up to a total $x$-derivative. Using this freedom we can choose $h$ in such a way that it satifsies a Riccati-type equation.

Indeed, from the
first two equations of the system (\ref{mateq}) written in componentwise form,
\begin{eqnarray*}
&&-\lambda {v_1}_x+ \,\frac12 {v_2}\,(2r- \lambda)
-\,\frac12 v\,(2 q-\lambda)=0 \\
&&-\lambda {v_2}_x+ \,v_1(2 q- \lambda)- 2 v_2 p=0\ ,
\end{eqnarray*}
we can write $v_1$ and $v_2$ as functions of $v=v_3$:
\begin{equation}
\label{eqv1v2}
\begin{array}{l}
v_1 =\displaystyle{\frac{-\lambda_1 v_x+2vp}{2r-\lambda}} \\
v_2 =\displaystyle{4 \, \frac{\lambda^2 r_x v_x-2 \lambda r_x p v}{(2r-\lambda)^3} +
2 \, \frac{- \lambda^2 v_{xx}+2\lambda p v_x+2\lambda v p_x}{(2r-\lambda)^2}
+ \frac{v( 2 q-\lambda)}{2r-\lambda}}\ .
\end{array}
\end{equation}
Substituting these expressions of $v_1$ and $v_2$ in (\ref{tr}) we obtain the following
 equation,
\begin{equation} \label{eqv}
\begin{array}{l}
(2r-\la)^3 F(\la)-{\la}^2(2r-\la)v_x^2+ 2{\la}^2 (2r-\la)v_{xx}v  -4 r_x\la^2\, v_x v\\
\qquad +v^2 \big( \la^3 +2 \la^2( 2p_x-q-2r) +4\la(-2p_x r+2r_x p+p^2+2qr+r^2)
\\ \qquad -8r(p^2+qr) \big)=0\ ,
\end{array}
\end{equation}
whose solution gives, with \rref{eqv1v2}, the differential $V(\lambda)$ of the Casimir $H(\lambda)$ 
of the Poisson pencil.

It is now easy to guess the most convenient choice of the total $x$-derivative
 involved in the definition of $h$. Putting
$$ h = \frac {v_x}{2v}\, + \, \frac{\sqrt{F(\la)}( 2r-\lambda)}{2\la v} $$
we get from \rref{eqv} the following Riccati equation for $h$:
\begin{eqnarray}\label{riccCH}
{} &(h_x+h^2)( 2\,r - \lambda ) {\lambda}^{2}-2 r_x \,h \, \lambda^2= -\frac14 {\lambda}^{3} \nonumber
+ ( \,r + \frac12 \,q -\,p_x  ) \,{\lambda}^{2}
\\{} &
+(-r  ^{2}- 2 \, r_x \, p  -2  \,q\,r +2\, p_x  r -p^2)\, \lambda+
 2\,q  r^{2}+2\,  p^{2}\,r\ .
\end{eqnarray}
If $h$ is a solution of this equation, then \rref{Hla} is a Casimir of the Poisson pencil and its coefficients constitute a bi-Hamiltonian hierarchy. We remark that equation \rref{riccCH}, evaluated at the points of $\CQ$, is the Riccati equation for the scalar CH hierarchy
(see \cite{Reyes} and \cite{clop}).

\section{The matrix CH hierarchy}

This section contains the main result of the paper, i.e., the existence of a hierarchy on 
$\CM = C^{\infty}(S^1,\mathfrak{sl}(2))$, to be called the matrix CH 
hierarchy, whose (bi-Hamiltonian) reduction coincides with the usual CH hierarchy.

We will to show that a suitable choice of a solution of the Riccati
equation \rref{riccCH} leads to a bi-Hamiltonian hierarchy which
starts from the Casimir \rref{casp2} of $P_2$ and contains the
3-component extension of the CH equation presented in the previous
section. To do this, we need to find a solution $h(\la)$ of (\ref{riccCH}) as a formal
series expansion in negative powers of $ \lambda$,
$$h=\sum_{i=0}^ {\infty} h_i \lambda^{-i}\ . $$
Substituting in (\ref{riccCH}) we obtain:
\begin{equation}\label{riccCHnl}
\begin{array}{l}
\big(2r \lambda^2-\lambda^3  \big) \Big(
\sum_{i=0}^ {\infty} \big( {h_{i}}_x+ \sum_{j=0}^{i}
h_{i-j}h_{j} \big) \lambda^{-i} \Big)
 -2r_x \lambda^2 \sum_{i=0}^{\infty} h_i \lambda^{-i} =\\ \noalign{\medskip}
=\, -\frac{\lambda^3}{4}
+\big( \frac{q}{2}+r-p_x \big)\lambda^2 - (+r^2+ 2rq -2r p_x +p^2+2r_x p) \lambda +2r(rq+p^2)
\end{array}
\end{equation}
It is possible to find the coefficients $h_i$ recursively, by
 solving a differential equation at every step. The first step corresponds to the coefficients of
$\lambda^3$ in \rref{riccCHnl}:
$$
\begin{array}{lc}
&{h_0}_x+h_0^2= \frac14\ .
\end{array}
$$
The only periodic solutions of this equation are $h_0=\pm\frac12 $. We choose the positive  solution, so that the next equation become
\begin{equation}
\label{h1}
{h_1}_x+h_1= -\frac12 (r+q)-r_x+p_x\ .
\end{equation}
The differential operator $\big(1+\partial_x \big)$
 is invertible in the space of periodic smooth functions
$C^{\infty}(S^1,\mathbb{R} )$,
so the previous equation has a unique solution
$$
h_1=\big(1+\partial_x \big)^{-1} m=
\int_{0}^{x}e^{y-x} m(y) \di y + \frac{1}{e-1} 
\int_{0}^{1} e^{y-x}  m(y) \di y\ ,
$$
were $m=-\frac12 (r+q)-r_x+p_x $. The next step is
\begin{equation*}
{h_2}_x+h_2= 2r({h_1}_x+h_1)
-h_1^2-2r_x h_1 +r^2+ 2rq -2r p_x +p^2+2r_x p \ .
\end{equation*}
Substituting the expression for $h_1$ we get
\begin{eqnarray*}
 h_2 &= \big(1+\partial_x \big)^{-1}
\Big[ 2rm - \big(\,\big(1+\partial_x \big)^{-1} m \big)^2-
 2r_x  \big(1+\partial_x \big)^{-1} m     \,
+r^2+ 2rq -2r p_x +p^2+2r_x p \Big]\\
&= \big(1+\partial_x \big)^{-1}
\Big[- \big(\,\big(1+\partial_x \big)^{-1} m \big)^2-
 2r_x  \big(1+\partial_x \big)^{-1} m     \,
+rq -2rr_x+p^2+2r_x p \Big]
\end{eqnarray*}
Similarly, we obtain
$$
\begin{array}{lc}
& {h_3}_x+h_3= 2r({h_2}_x+h_2+ h_1^2)-2 h_1h_2-2 r_x h_2 -2r(qr+p^2) \ ,
\end{array}
$$
so that
$$
\begin{array}{l}
h_3= \big(1+\partial_x \big)^{-1}\times\\
\Big [ 2r \big (2rm- \big(\, \big(1+\partial_x \big)^{-1}m\big)^2
-2r_x \big(1+\partial_x \big)^{-1}m +r^2+ 2rq -2r p_x +p^2+2r_x p \big) \\+2r
 \big(\,\big(1+\partial_x \big)^{-1} m \big)^2 -
  \big(\, 2r_x+ \big(1+\partial_x \big)^{-1} m \big)
  \big(\, \big(1+\partial_x \big)^{-1}
\big( 2rm - \big(\,\big(1+\partial_x \big)^{-1} m \big)^2\\
  - 2r_x  \big(1+\partial_x \big)^{-1} m\,+r^2
+ 2rq -2r p_x +p^2+2r_x p \big) \big)
  -2r(qr+p^2)\Big]
\end{array}
$$
and so on. 
The {\em matrix CH hierarchy\/} is the bi-Hamiltonian hierarchy on $(\CM,P_1,P_2)$ given by the functionals
$H_j=\int h_j \di x$, for $j\ge 1$. This corresponds to putting $H(\lambda)=\sum_{j\ge 1}H_j\lambda^{-j}$
and $F(\lambda)=\frac{1}{4\lambda^2}$ in equation \rref{Hla}.

Integrating both sides of (\ref{h1}) with respect to $x$ we obtain
\begin{equation*}
\int h_1 \di x=-\frac{1}{2}\int (r+q)\di x= H_1\ .
\end{equation*}
Moreover, using the identity
\begin{equation*}
m=\frac{1}{2}(r-q)-(r+r_x)+p_x=(1-\partial_x^2)\gamma-(1+\partial_x)r\ ,
\end{equation*}
that is,
\begin{equation*}
(1+\partial_x)^{-1}m=\gamma-\gamma_x-r\ ,
\end{equation*}
it is easy to show that
\begin{equation*}
\int h_2 \di x=\int(p^2-\gamma^2-\gamma_x^2) \di x
\end{equation*}
coincides with the first Hamiltonian $H_2$ of the vector field \rref{nlch2}.

\section{Conclusions}

In this paper we showed that the Camassa-Holm equation has a property that seems to be a common feature of integrable PDEs, namely, that it can be interpreted as a reduction of a member of a matrix hierarchy. Indeed, we constructed a bi-Hamiltonian hierarchy for the Poisson pair $(P_1=[\,\cdot\, ,S],P_2=\partial_x  +[\,\cdot\, , A])$, starting from a Casimir of $P_2$ and giving rise to the scalar CH hierarchy after a (bi-Hamiltonian) reduction. In particular, this allowed us to find a 3-component extension of the Camassa-Holm equation. We conclude with two natural developments of our study.

First of all, the restriction of the matrix CH hierarchy to the symplectic leaf $\CS$, defined by the constraint $p^2+qr=0$, is by construction a 2-component extension of the CH hierarchy. Is there a parametrization of $\CS$ showing that this extension 
coincides with the (bi-Hamiltonian) one discussed in \cite{CLZ,F,LZ}? If the answer is affirmative, how can the bi-Hamiltonian structure on $\CM$ be reduced on $\CS$?

Secondly, the scalar CH hierarchy has a negative (or local) counterpart, generated by a Casimir 
of $\left(P_1^{\mbox{\scriptsize rd}}\right)_{q} =2(2q\partial_x+q_x)$. In order to find a local matrix CH hierarchy, one should start from the Casimir $H_0=p^2+qr$ of $P_1$. Since $H_0$ vanishes on the symplectic leaf $\CS$ used in the reduction procedure, the construction of an extension of the local CH hierarchy is more complicated and requires a  careful description from the geometric point of view.

\end{document}